# Multi-Bernoulli Sensor-Control via Minimization of Expected Estimation Errors

Amirali K. Gostar, Reza Hoseinnezhad and Alireza Bab-Hadiashar

*Abstract*—This paper presents a sensor-control method for choosing the best next state of the sensor(s), that provide(s) accurate estimation results in a multi-target tracking application. The proposed solution is formulated for a multi-Bernoulli filter and works via minimization of a new estimation error-based cost function. Simulation results demonstrate that the proposed method can outperform the state-of-the-art methods in terms of computation time and robustness to clutter while delivering similar accuracy.

## I. INTRODUCTION

SENSOR-control techniques in multi-target tracking problems are designed to find the optimal control command from a set of *admissible* commands that results in the *most rewarding* observations. Sensor-control techniques usually involve addressing two main components:
- multi-object filtering, and
- sequential decision making.

To solve the above problems, we formulate the sensor-control problem in the *partially observed Markov decision process* (POMDP) framework by proposing to use a multi-Bernoulli filter for the multi-object filtering task. This filter enables us to devise a new cost function which has a more direct relation with the OSPA [36] multi-target miss-distance compared to information-theoretic-based objective functions commonly used in sensor-control solutions. Consequently, it results in better performance (in terms of the OSPA error).

The purpose of multi-object filtering is to jointly estimate the randomly varying number and states of multiple objects. Several solutions for this problem have recently been developed in the finite set statistics (FISST) framework, in which the multi-object entity is treated as a random finite set (RFS) with its distribution being predicted and updated in every time step [29]. Examples of such solutions include the PHD filter [25], [29], CPHD filter [28], [42], MeMBer filter [29], cardinality balanced MeMBer (CB-MeMBer) filter [43] and Generalized labeled multi-Bernoulli filter [41].

A number of sensor-control solutions have also been developed to work within the FISST multi-target filtering schemes [9]–[12], [24], [26], [30], [34], [35], [45], [47]. In these solutions, a criterion is defined to evaluate the *expected quality* of the updated multi-target density after an admissible command is applied to the sensor. In this approach, the chosen control command is the one that provides the *best* updated density in terms of the defined criterion. Examples of this type of criteria include the Csiszár information-theoretic functional [27], posterior expected number of targets (PENT) [30], [47] and the Cauchy-Schwarz divergence between the probability densities of two Poisson point processes [13].

The most common criterion to evaluate the quality of the updated distribution is its *divergence* from the predicted distribution [34]. In two consecutive papers, Ristic et al. [34], [35] used Rényi divergence as a reward function to quantify the information gained via updating the predicted density in a random set filtering framework. In [34], the implementation of Rényi reward maximization was investigated for the "general form" of multi-target filters. Since this approach is computationally intractable even for a small number of targets, in the second paper [35], the Rényi divergence function was approximated based on i.i.d. cluster assumption for the multi-target distribution within a PHD filtering scheme. Later, in [12] and [14] the Rényi divergence objective function was implemented via the multi-Bernoulli filter.

In our recent study [9], reported independently from [12] and [14], we introduced the use of multi-Bernoulli filtering in a task-driven objective function. This criterion is introduced in the form of a *cost* that would be minimized over all admissible control commands. It is important to note that in multi-Bernoulli filter, estimation of the number and states of targets from the updated multi-Bernoulli distribution is straightforward and no clustering is needed. Furthermore, multi-Bernoulli filters have shown to generally provide effective solutions for multi-target tracking methods in various multi-target applications using radar, video and audio-visual measurements [15]–[18], [40], [43]. We also note that both in our earlier sensor-control method, and in the method introduced in this paper, the cardinality-balanced multi-Bernoulli filter (CB-MeMBer) [43] is used as an appropriate multi-object filtering scheme for the sensor-control method. CB-MeMBer is proven to result in *unbiased* cardinality estimates under high SNR [43] .

In our previous work [9], the cost function was defined in terms of the estimation error of *only* the cardinality of the multi-target state with the assumption that in practice, an accurate estimate of the number of target would be followed by accurate estimates of the actual target states. Around the same time, a similar approach based on using a variance-based cost function was also pursued by Hoang et al. [14]. Unlike the cost function presented in [9], in which the statistical variance of cardinality around its *mean* is chosen as the cost function, Hoang et al. [14] used the variance of cardinality around its *MAP estimate* as the cost function.

All authors are with School of Aerospace, Mechanical and Manufacturing Engineering, RMIT University, Victoria 3083, Australia.

This work was supported by the ARC Discovery Project grant DP130104404.

In this paper, we present a new cost function called *Posterior Expected Error of Cardinality and States* (PEECS) that its minimization can lead to a high quality updated density even with limited observability. The proposed cost function is intended to account for both localization and cardinality errors. Inclusion of the extra terms for localization errors in PEECS (compared to cardinality variance-only costs used in [9], [12]) can enhance the performance of sensor-control, especially in challenging cases with numerous targets and moderate to high rates of clutter.

The rest of paper is organized as follows. In Sec. II, a formal statement of the sensor-control problem is presented. Since, our sensor-control solution is specifically formulated to work within multi-Bernoulli filtering schemes, the cardinality-balanced multi-Bernoulli filter is briefly reviewed in Sec. III. Then, the details of sensor-control method are presented in Sec. IV followed by comparative numerical studies in Sec. VI. Sec. VII concludes the paper.

## II. PROBLEM STATEMENT: SENSOR-CONTROL

This paper focuses on developing an effective *sensor-control* solution for applications where *mobile* sensors are used for multi-object tracking while only instantaneous performance is considered (myopic approach).

The main aim of sensor-control is to choose the *best* control command(s) from a number of *admissible* commands. The application of the chosen command changes the sensor(s) state e.g. the location, heading or both to reach more accurate estimation of the number and states of targets via a set of noisy measurements which may include by clutter. The quality of a sensor-control command is usually evaluated in terms of a reward function (to be maximized) or a cost function (to be minimized). Following [3], [34], [35], we formulated the sensor-control problem in the *Partially Observable Markov Decision Process* (POMDP) framework. POMDP is a generalized form of Markov Decision Process (MDP) [19] in which there is no direct access to the *states* and the states information are only realized by noisy observations. The POMDP framework employed in this paper to formulate the sensor-control problem includes the following elements at any time $k$:

- a finite set $X_k$ comprising single-object states;
- a set of sensor-control commands (actions) $\mathbb{U}$;
- a stochastic model for single-target state transition $f_{k|k-1}(x_k|x_{k-1})$;
- a finite set of observations $Z$, generally made by $N$ sensors[1];
- a stochastic measurement model $g_k(z|x)$; and
- a function $\mathcal{V}(u;X_k)$ that returns a reward or cost associated with action command $u \in \mathbb{U}$.

In principle in a POMDP framework, the action space is infinite and continuous. To reduce computational complexity, we assume that the sensor(s) can choose from a finite set of actions or commands, namely *admissible control commands* [1].

[1]Without loss of generality, in our formulations and simulation studies, we consider multiple observations from one sensor and note that extension of the proposed solution to the general multi-sensor case is straightforward.

The purpose of sensor-control is to find the control command $\mathring{u}$ which optimizes the cost (or reward) function. In stochastic filtering, where the multi-target states $X_{k-1}$ and $X_k$ are characterized by their distributions, the control command $\mathring{u}$ is commonly chosen to optimize the statistical mean of the objective function $\mathcal{V}(u;X_k)$ over all observations. The function $\mathcal{V}(u;X_k)$ is a real-valued objective function of the control command and estimated state of the targets. Depending on the nature of this function (being either a "*cost*" or a "*reward*"), the sensor-control solution would *minimize* or *maximize* it,

$$\mathring{u}_k = \underset{u\in\mathbb{U}}{\operatorname{argmin}} / \underset{u\in\mathbb{U}}{\operatorname{argmax}} \left\{ \mathbb{E}_{Z_k(u)}[\mathcal{V}(u;X_k)] \right\}. \qquad (1)$$

The objective function usually depends on the statistical distribution of $X_{k-1}$ and $X_k$, which can be recursively computed in a Bayesian filtering scheme. The latest development in multi-target filtering is Mahler's *finite set statistics* (FISST) framework which provides a set of mathematical tools that allows direct application of Bayesian inferencing to multi-target problems. In a seminal paper [39], Vo et al. established the relationship between FISST and conventional probability. The theory provides a rigorous paradigm for calculations involving RFSs based upon the notion of integration and density in *point process* and *stochastic geometry* [6], [31], [37]. This approach to multi-target filtering has influenced a variety of research areas such as autonomous vehicles and robotics [32], radar tracking [21] and image processing [23].

Since the general form of the multi-target Bayes filter is intractable [29], the FISST framework also includes a number of computationally feasible approximations. Examples of such filters include the probability hypothesis density (PHD) [25], the cardinalized PHD (CPHD) [28], the multi-Bernoulli (MeMBer) [29] and the cardinality balanced MeMBer (CB-MeMBer) [43] filters. PHD and CPHD filters are implemented via *Sequential Monte Carlo* (SMC) method [39], [44], [46] or using Gaussian mixtures [38], [42]. The convergence of PHD and CPHD filters has been studied in [2], [4], [5], [39]. SMC and Gaussian mixture implementations of CB-MeMBer filter have also appeared in [43].

As mentioned earlier, due to the better accuracy and lower computational requirements of multi-Bernoulli filtering schemes, in applications where SMC is necessary, this paper focuses on an effective sensor-control solution for CB-MeMBer filter. The common approach to sensor-control within multi-target filtering schemes is to choose the reward/cost function in terms of the predicted multi-target state and the expected update outcomes for every admissible control command. Before we present our choice of cost function, the multi-Bernoulli filtering scheme is briefly reviewed in the next section.

## III. CB-MeMBer FILTER

A multi-Bernoulli RFS is defined as the union of $M$ independent Bernoulli RFSs $X^{(i)}, i = 1, \ldots, M$. Each Bernoulli RFS is characterized by the parameter $r^{(i)}$ and function $p^{(i)}(\cdot)$, where $r^{(i)}$ is the probability of existence of an element in $X^{(i)}$ (being a singleton) and $p^{(i)}(\cdot)$ is the probability density function (pdf) of the state of the single element of the set if it exists. It has been shown that all statistical properties of a multi-Bernoulli RFS can be completely characterized by the set of $M$ parameter pairs, existence probabilities $r^{(i)}$ and the distributions $p^{(i)}(\cdot)$, and a general multi-Bernoulli RFS is commonly denoted by $X \sim \{(r^{(i)}, p^{(i)})\}_{i=1}^{M}$ [29].

The multi-Bernoulli Bayesian filter was first derived by Mahler [29]. He showed that if the prior distribution of the multi-target random set state is multi-Bernoulli, the predicted and updated posteriors are also approximately multi-Bernoulli. Later, the original filter was shown to produce biased cardinality estimates [43] and a modified version, called Cardinality-Balanced MeMBer (CB-MeMBer) filter, was formulated in a numerically tractable form and implemented by the Sequential Monte Carlo (SMC) [43].

Suppose that at time $k-1$, the posterior multi-Bernoulli density of the multi-target state is given by $\{(r_{k-1}^{(i)}, p_{k-1}^{(i)})\}_{i=1}^{M_{k-1}}$ and each $p_{k-1}^{(i)}$ is approximated by a set of particles:

$$p_{k-1}^{(i)}(x) = \sum_{j=1}^{L_{k-1}^{(i)}} w_{(k-1)}^{(i,j)} \delta_{x_{k-1}^{(i,j)}}(x). \quad (2)$$

For existing and new born targets, two proposal densities are given and denoted by $q_k^{(i)}(\cdot|x_{k-1}, Z_k)$ and $b_k^{(i)}(\cdot|Z_k)$. The predicted multi-target density (also a multi-Bernoulli density) is represented by the union of the survived targets and the newly born targets:

$$\pi_{k|k-1} = \left\{(r_{P,k|k-1}^{(i)}, p_{P,k|k-1}^{(i)})\right\}_{i=1}^{M_{k-1}} \cup \left\{(r_{\Gamma,k}^{(i)}, p_{\Gamma,k}^{(i)})\right\}_{i=1}^{M_{\Gamma,k}} \quad (3)$$

where existence probabilities and distributions of the predicted Bernoulli components are given by:

$$\begin{aligned}
r_{P,k|k-1}^{(i)} &= r_{k-1}^{(i)} \sum_{j=1}^{L_{k-1}^{(i)}} w_{k-1}^{(i,j)} \, p_{S,k}(x_{k-1}^{(i,j)}) \\
p_{P,k|k-1}^{(i)}(x) &= \sum_{j=1}^{L_{k-1}^{(i)}} \tilde{w}_{P,k|k-1}^{(i,j)} \, \delta_{x_{P,k|k-1}^{(i,j)}}(x) \\
r_{\Gamma,k}^{(i)} &= \text{birth model parameters} \\
p_{\Gamma,k}^{(i)}(x) &= \sum_{j=1}^{L_{\Gamma,k}^{(i)}} \tilde{w}_{\Gamma,k}^{(i,j)} \, \delta_{x_{\Gamma,k}^{(i,j)}}(x)
\end{aligned} \quad (4)$$

where

$$\begin{aligned}
x_{P,k|k-1}^{(i,j)} &\sim q_k^{(i)}(\cdot|x_{k-1}^{(i,j)}, Z_k), j = 1, \cdots, L_{k-1}^{(i)} \\
w_{P,k|k-1}^{(i,j)} &= \frac{w_{k-1}^{(i,j)} f_{k|k-1}(x_{P,k|k-1}^{(i,j)}|x_{k-1}^{(i,j)}) \, p_{S,k}(x_{k-1}^{(i,j)})}{q_k^{(i)}(x_{P,k|k-1}^{(i,j)}|x_{k-1}^{(i,j)}, Z_k)} \\
\tilde{w}_{P,k|k-1}^{(i,j)} &= w_{P,k|k-1}^{(i,j)} / \sum_{\ell=1}^{L_{k-1}^{(i)}} w_{P,k|k-1}^{(i,\ell)} \\
x_{\Gamma,k}^{(i,j)} &\sim b_k^{(i)}(\cdot|Z_k), j = 1, \cdots, L_{\Gamma,k}^{(i)} \\
w_{\Gamma,k}^{(i,j)} &= p_{\Gamma,k}(x_{\Gamma,k}^{(i,j)}) / b_k^{(i)}(x_{\Gamma,k}^{(i,j)}|Z_k) \\
\tilde{w}_{\Gamma,k}^{(i,j)} &= w_{\Gamma,k}^{(i,j)} / \sum_{\ell=1}^{L_{\Gamma,k}^{(i)}} w_{\Gamma,k}^{(i,\ell)}.
\end{aligned} \quad (5)$$

Let us denote the predicted multi-Bernoulli distribution by $\{(r_{k|k-1}^{(i)}, p_{k|k-1}^{(i)})\}_{i=1}^{M_{k|k-1}}$. The updated multi-Bernoulli is represented by the union of *legacy tracks* and *measurement-corrected tracks* [29], [43]:

$$\pi_k = \left\{(r_{L,k}^{(i)}, p_{L,k}^{(i)})\right\}_{i=1}^{M_{k|k-1}} \cup \left\{(\overset{\star}{r}_{U,k}(z), \overset{\star}{p}_{U,k}(\cdot;z))\right\}_{z \in Z_k} \quad (6)$$

with the following existence probabilities and singleton densities:

$$\begin{aligned}
r_{L,k}^{(i)} &= r_{k|k-1}^{(i)} \frac{(1-\varrho_{L,k}^{(i)})}{(1-r_{k|k-1}^{(i)} \varrho_{L,k}^{(i)})} \\
p_{L,k}^{(i)}(x) &= \sum_{j=1}^{L_{k|k-1}^{(i)}} \tilde{w}_{L,k}^{(i,j)} \delta_{x_{k|k-1}^{(i,j)}}(x) \\
\overset{\star}{r}_{U,k}(z) &= \frac{\sum_{i=1}^{M_{k|k-1}} \frac{r_{k|k-1}^{(i)}(1-r_{k|k-1}^{(i)})\varrho_{U,k}^{(i)}(z)}{(1-r_{k|k-1}^{(i)} \varrho_{L,k}^{(i)})^2}}{\kappa_k(z) + \sum_{i=1}^{M_{k|k-1}} \frac{r_{k|k-1}^{(i)} \varrho_{U,k}^{(i)}(z)}{1 - r_{k|k-1}^{(i)} \varrho_{L,k}^{(i)}}} \\
\overset{\star}{p}_{U,k}(x;z) &= \sum_{i=1}^{M_{k|k-1}} \sum_{j=1}^{L_{k|k-1}^{(i)}} \tilde{w}_{U,k}^{\star(i,j)}(z) \delta_{x_{k|k-1}^{(i,j)}}(x)
\end{aligned} \quad (7)$$

where

$$\begin{aligned}
\varrho_{L,k}^{(i)} &= \sum_{j=1}^{L_{k|k-1}^{(i)}} w_{k|k-1}^{(i,j)} p_{D,k}(x_{k|k-1}^{(i,j)}) \\
\tilde{w}_{L,k}^{(i,j)} &= w_{L,k}^{(i,j)} / \sum_{j=1}^{L_{k|k-1}^{(i)}} w_{L,k}^{(i,j)} \\
w_{L,k}^{(i,j)} &= w_{k|k-1}^{(i,j)}(1 - p_{D,k}(x_{k|k-1}^{(i,j)})) \\
\varrho_{U,k}^{(i)}(z) &= \sum_{j=1}^{L_{k|k-1}^{(i)}} w_{k|k-1}^{(i,j)} p_{D,k}(x_{k|k-1}^{(i,j)}) g_k(z|x_{k|k-1}^{(i,j)}) \\
\tilde{w}_{U,k}^{\star(i,j)}(z) &= \overset{\star}{w}_{U,k}^{(i,j)}(z) / \sum_{i=1}^{M_{k|k-1}} \sum_{j=1}^{L_{k|k-1}^{(i)}} \overset{\star}{w}_{U,k}^{(i,j)}(z) \\
\overset{\star}{w}_{U,k}^{(i,j)}(z) &= \frac{w_{k|k-1}^{(i,j)} r_{k|k-1}^{(i)} p_{D,k}(x_{k|k-1}^{(i,j)}) g_k(z|x_{k|k-1}^{(i,j)})}{1 - r_{k|k-1}^{(i)}}.
\end{aligned} \quad (8)$$

To avoid degeneracy, the CB-MeMBer filter also includes a resampling step for each track. To prevent the numerical explosion of the generated particles, the existence probabilities are thresholded and the low probability tracks are removed while similar ones are merged – details of these steps are provided in [43].

## IV. Multi-Bernoulli sensor-control

In a multi-Bernoulli target tracking scheme, assume that at time $k-1$, the multi-target distribution parameters are $\{r_{k-1}^{(i)}, p_{k-1}^{(i)}\}_{i=1}^{M_{k-1}}$ where $M_{k-1}$ indicates maximum number of targets. This distribution is propagated through the multi-Bernoulli prediction and update steps, and turns into an updated multi-Bernoulli distribution with parameters $\{r_k^{(i)}, p_k^{(i)}\}_{i=1}^{M_k}$. As part of the multi-target tracking solution, at each time $k$, the number and states of targets are extracted from the updated multi-Bernoulli parameters. We note that the sensor measurements are used within the update step of the recursion, and affect the quality of the updated distribution.

In practice, the quality of sensor measurements usually depends on a sensor state (e.g. the sensor location) which is assumed to be controllable, and the sensor-control problem is focused on choosing the command that would lead to the best sensor state. As it was mentioned earlier, most solutions are based on maximizing an information theoretic reward function such as Rényi divergence [20], [34], [35]. The main rationale behind choosing such reward functions is that the information encapsulated by the estimated multi-target distribution is expected to gradually increase as further measurements become available in time.

In this paper, we take a different approach in which the updated multi-Bernoulli distribution parameters are *directly* utilized to define a new cost function. Our approach is to define a cost function that quantifies the average uncertainty in all possible multi-target state estimates after each update step.[2] The main difference here is that our focus is on measuring the quality of the updated density in terms of level of uncertainties, not the information gained from prediction to update (e.g. Rényi divergence function).

The updated distribution depends on the measurement set which is a function of the chosen sensor-control command. In principle, the whole distribution of all possible measurement sets is used to compute the update distribution. However, using whole measurement distribution is computationally expensive if not intractable. Thus, to reduce the computational complexity, for each sensor control command, only *the predicted ideal measurement set* (PIMS) is used [27]. In PIMS approach only one future measurement sample set is generated for each control command under the ideal conditions of no measurement noise and no clutter where probability of detection is equal to one.[3] In order to define the new cost function, we note that the PIMS depends on the chosen control command. For each command, we first compute the PIMS, then calculate an updated multi-object distribution $\{r_k^{(i)}, p_k^{(i)}\}_{i=1}^{M_k}$ by considering the PIMS as the acquired measurement.

A linear combination of the normalized errors of the number of targets and their estimated states are considered as a measure of uncertainty associated with estimation of the multi-target state and as the cost function:

$$\mathcal{V}(u; X_k) = \eta\ \varepsilon_{|X|}^2(u) + (1 - \eta)\ \varepsilon_X^2(u), \quad (9)$$

where $\varepsilon_{|X|}^2(u)$ denotes the normalized error of estimated cardinality of the multi-target state and $\varepsilon_X^2(u)$ denotes the normalized error of the multi-target state estimate. The details of defining and computing the normalized error terms, $\varepsilon_{|X|}^2(u)$ and $\varepsilon_X^2(u)$ for SMC implementation are presented in Sec. V through equations (10)–(16). Note that $\eta \in [0, 1]$ is a user-defined constant parameter to tune the influence of the error terms on the total sensor control cost. It is also important to note that the expectation term in (1) does not appear as we use the PIMS approach [27] instead of sampling and averaging in measurement space.

## V. Implementation

At time $k-1$, the multi-target distribution is modeled by a multi-Bernoulli RFS with parameters $\{r_{k-1}^{(i)}, p_{k-1}^{(i)}\}_{i=1}^{M_{k-1}}$, where $M_{k-1}$ indicates the maximum possible number of targets, and each $p_{k-1}^{(i)}$ for $i = 1, \ldots, M_{k-1}$ is approximated by $L_{k-1}^{(i)}$ particles denoted by $\{w_{k-1}^{(i,j)}, x_{k-1}^{(i,j)}\}_{j=1}^{L_{k-1}^{(i)}}$. In the prediction step, the multi-Bernoulli filter propagates the multi-Bernoulli components based on the temporal information from the transition density ($f_{k|k-1}(\cdot)$), the probability of survival ($P_s$), and the predefined multi-Bernoulli birth terms [43]. The predicted multi-Bernoulli density is denoted by $\{r_{k|k-1}^{(i)}, p_{k|k-1}^{(i)}\}_{i=1}^{M_{k|k-1}}$ where $p_{k|k-1}^{(i)}$ is approximated by $L_{k|k-1}^{(i)}$ particles and represented by $\{w_{k|k-1}^{(i,j)}, x_{k|k-1}^{(i,j)}\}_{j=1}^{L_{k|k-1}^{(i)}}$ for $i = 1, \ldots, M_{k|k-1}$.

As it was mentioned before, the updated distribution would depend on the measurement set which in turn would be dependent on the chosen sensor control command. Similar to Ristic et al. [33], for each control command, a single pseudo-measurement is generated for each single object under ideal conditions (perfect detection, zero measurement noise and no clutter). The ensemble of such pseudo-measurements forms the PIMS. To compute each of the ideal measurements, we need to know the single-target states which can be estimated from the predicted multi-target distribution via the following two steps.[4] First, the estimated number of targets, denoted by $\hat{M}_{k|k-1}$ is obtained by counting the predicted existence probabilities that exceed a given threshold (e.g. 0.5 as an unbiased value). Then, the EAP estimate of the $\ell$-th target state is given by $\hat{x}_{k|k-1}^{(\ell)} = \sum_{j=1}^{L_{k|k-1}^{i_\ell}} w_{k|k-1}^{(i_\ell,j)} x_{k|k-1}^{(i_\ell,j)}$ where the indices $i_\ell$ refer to the Bernoulli components that were counted in step I.

Having the ideal measurement set for each admissible sensor command, $u_k \in \mathbb{U}_k$, and considering it as the actual measurement, we can then run the update step and calculate the cost corresponding to that command. The cost defined in (9) combines two normalized error terms, $\varepsilon_{|X|}^2(u)$ for the cardinality estimate and $\varepsilon_X^2(u)$ for the multi-target state estimate. Both terms depend on the updated multi-object posterior $\{r_{k,u_k}^{(i)}, p_{k,u_k}^{(i)}(\cdot)\}_{i=1}^{M_{k,u_k}}$ which in turn depends on the PIMS computed for the command $u_k$.

In [7] and [8], Delande et al. showed that the regional variance of the number of targets quantifies the *certainty* of the filter estimates of the number of the targets that evolve in surveillance region. Following [8], we chose the variance of the cardinality as a meaningful measure for its uncertainty or estimation error. In terms of the updated probabilities of existence this variance is given by:

$$\sigma_{|X|}^2(u_k) = \sum_{i=1}^{M_{k,u_k}} \left[ r_{k,u_k}^{(i)} (1 - r_{k,u_k}^{(i)}) \right]. \quad (10)$$

It is important to note that in the linear combination presented in equation (9), in order for the weight $\eta$ to be dimensionless and bounded within the $[0, 1]$ interval (so that both $\eta$ and $1-\eta$ weights are positive), the two error terms have to be dimensionless. We choose "normalized" error terms which are both dimensionless and bounded within $[0,1]$ themselves. Thus, the linear combination forming the cost in (9) will be guaranteed to be bounded within $[0,1]$. The normalized cardinality error term can be computed as follows:

$$\varepsilon_{|X|}^2(u_k) = \frac{\sigma_{|X|}^2(u_k)}{\max\{\sigma_{|X|}^2(u_k)\}}. \quad (11)$$

---

[2] It is important to note that this cost is not totally independent of the prediction outcomes, since state estimates extracted from predicted multi-Bernoulli density are used to calculate the proposed cost function.

[3] In situations where the probability of detection is high and clutter rate is moderate, the multi target posterior density can be approximated under the assumptions of the PIMS (see [27]–section 4.1 and [33]).

[4] It is important to note that same procedure is used to finally extract the multi-target estimates after the control command is chosen and applied, the measurements are acquired from sensors and the multi-target state is updated. Furthermore, we note that alternative methods (such as the method suggested in [43]) could be also used for this purpose.

The maximum variance occurs when $\forall i, r_{k,u_k}^{(i)} = 0.5$.

$$\max\{\sigma_{|X|}^2(u_k)\} = \frac{M_{k,u_k}}{4}.$$

To arrive at a meaningful measure for the normalized state estimation error term $\varepsilon_X^2(u)$ in the cost defined in (9), we first compute the following total state estimation error:

$$\varepsilon_X^2(u_k) = \sum_{i=1}^{M_{k,u_k}} \left[ r_{k,u_k}^{(i)} \varsigma_{x^{(i)}}^2(u_k) \right] \bigg/ \sum_{i=1}^{M_{k,u_k}} r_{k,u_k}^{(i)} \quad (12)$$

which is the weighted average of estimation errors of the states of single targets associated with each single Bernoulli component. The rationale behind this choice of weights is that Bernoulli components with larger probabilities of existence contribute more strongly to the EAP estimate of the multi-object state –see section IV-A.4 in [43]. Although $\varepsilon_X^2$ is not directly related to estimation error, we named it as the "total state estimation error". The philosophy behind this choice lies in the fact that reduction of $\varsigma_{x^{(i)}}^2$ increases the *confidence of estimation* which implicitly increases the expected accuracy of estimation.

To compute the single Bernoulli component errors $\varsigma_{x^{(i)}}^2(u_k)$, we note that in practice, minimizing the estimation error of only some *selected elements* of target states maybe of the interest. For instance, in target tracking, usually the prime interest is in *location*, and target speed is included in the single-target state vector due to its appearance in motion and perhaps measurement models. In such target-tracking applications, an intuitive scalar measure for the single Bernoulli component error is given by the product of the variances of the target location coordinates. If the stochastic variations of target location coordinates are independent, this measure will translate into the absolute determinant of the covariance matrix of the target location. In case of tracking multiple-targets in two–dimensional space, the single Bernoulli component error term, $\varsigma_{x^{(i)}}^2(u_k)$, is given by:

$$\varsigma_{x^{(i)}}^2(u_k) \propto \sigma_{x^{(i)}}^2(u_k) \; \sigma_{y^{(i)}}^2(u_k) \quad (13)$$

where x and y denote the x and y-coordinates of the single-target location (part of its state vector $x$), and the proportionality (not equality) is chosen because the above product does not lead to a normalized measure. We note that in theory, a very small $\sigma_{x^{(i)}}^2(u_k)$ and an unacceptably large $\sigma_{y^{(i)}}^2(u_k)$ can still lead to a relatively small $\varsigma_{x^{(i)}}^2(u_k)$. However, a near zero determinant of covariance matrix for target locations would contradict our independence assumption for the target movements. A remedy would be to take the sum of normalized variances along the coordinates. But unlike the product, this sum would have no meaningful interpretation. Having the updated particles and weights of each Bernoulli component, the single-coordinate errors can be calculated as follows:

$$\sigma_{x^{(i)}}^2(u_k) = \sum_{j=1}^{L_{k,u_k}^{(i)}} w_{k,u_k}^{(i,j)} x_{k,u_k}^{(i,j)2} - \left(\sum_{j=1}^{L_{k,u_k}^{(i)}} w_{k,u_k}^{(i,j)} x_{k,u_k}^{(i,j)}\right)^2$$

$$\sigma_{y^{(i)}}^2(u_k) = \sum_{j=1}^{L_{k,u_k}^{(i)}} w_{k,u_k}^{(i,j)} y_{k,u_k}^{(i,j)2} - \left(\sum_{j=1}^{L_{k,u_k}^{(i)}} w_{k,u_k}^{(i,j)} y_{k,u_k}^{(i,j)}\right)^2$$
(14)

where $x_{k,u_k}^{(i,j)}$ and $y_{k,u_k}^{(i,j)}$ denote the coordinates extracted from the particle $x_{k,u_k}^{(i,j)}$ and power operation is element-wise. To normalize the total state estimation error term $\varsigma_{x^{(i)}}^2(u_k)$ in (13), we note that with equally weighted particles, i.e. when $\forall j, w_{k,u_k}^{(i,j)} = 1/L_{k,u_k}^{(i)}$, the particles representing the $i$-th single Bernoulli component do not convey any information and the above estimation variances adopt their maximum values as follows:

$$\max\{\sigma_{x^{(i)}}^2(u_k)\} = \frac{1}{L_{k,u_k}^{(i)}}(1 - \frac{1}{L_{k,u_k}^{(i)}}) \sum_{j=1}^{L_{k,u_k}^{(i)}} x_{k,u_k}^{(i,j)2}$$
$$\max\{\sigma_{y^{(i)}}^2(u_k)\} = \frac{1}{L_{k,u_k}^{(i)}}(1 - \frac{1}{L_{k,u_k}^{(i)}}) \sum_{j=1}^{L_{k,u_k}^{(i)}} y_{k,u_k}^{(i,j)2}.$$
(15)

The single Bernoulli error terms $\varsigma_{x^{(i)}}^2(u_k)$ in (13) can be normalized as follows:

$$\varsigma_{x^{(i)}}^2(u_k) = \frac{\sigma_{x^{(i)}}^2(u_k) \; \sigma_{y^{(i)}}^2(u_k)}{\max\{\sigma_{x^{(i)}}^2(u_k)\} \; \max\{\sigma_{y^{(i)}}^2(u_k)\}} \quad (16)$$

and the computed values are then used in (12) to calculate the normalized state estimation error term in the cost. We note that the product of variance terms in (13) is a squared variance, but after normalization, it would be still a normalized measure for confidence of estimation. Extension of the terms derived in (13)–(16) to the cases involving more than two dimensional location parameters is straightforward.

Having the cost values computed for all admissible sensor control commands, the best command $\overset{\star}{u}_k$ is then chosen as the one incurring the smallest cost:

$$\overset{\star}{u}_k = \underset{u_k \in \mathbb{U}_k}{\arg\min} \, \mathcal{V}(u_k; X_{k-1}). \quad (17)$$

Algorithm 1 shows the pseudocode for the CB-MeMBer multi-target filtering scheme with the proposed sensor-control.

## VI. SIMULATION RESULTS

The performance of the proposed PEECS sensor-control method was evaluated in two challenging case studies. Case study 1 is borrowed from [35] and [14] and used to compare the performance of PEECS with the PHD-based sensor control method for different clutter rates. Case study 2 is a more challenging extension of case study 1 in which the targets are not pseudo stationary (they move and turn through the scenario), the motion model is more complex, and in addition to range, bearing measurements are also available.

**Algorithm 1** The CB-MeMBer multi-target filtering recursion with PEECS sensor-control.

INPUTS: time $k$, dynamic model $f_{k|k-1}(\cdot|x_{k-1})$, multi-Bernoulli birth model parameters, prior multi-Bernoulli parameters from time $k-1$, detection probability $p_D(\cdot)$, measurement likelihood function $g_k(\cdot|x)$, and clutter intensity $\nu(\cdot)$ and its integral $\lambda_c$, current sensor(s) location(s), finite set of admissible sensor-control commands $\mathbb{U}$.

OUTPUT: The best control command $\overset{\star}{u}$ and updated multi-Bernoulli parameters.

**Prediction:**
1: Using equations (1)–(3), compute the predicted multi-Bernoulli component parameters and their particles $\{r_{k|k-1}^{(i)}, \{w_{k|k-1}^{(i,j)}, x_{k|k-1}^{(i,j)}\}_{j=1}^{L_{k|k-1}^{(i)}}\}_{i=1}^{M_{k|k-1}}$.

**Pre-estimation:**
2: $\ell \leftarrow 0$,
3: **for** $i = 0, M_{k|k-1}$ **do**
4:    **if** $r_{k|k-1}^{(i)} > 0.5$ **then**
5:       $\ell \leftarrow \ell + 1$
6:       $\hat{x}_{k|k-1}^{(\ell)} = \sum_{j=1}^{L_{k|k-1}^{(i)}} w_{k|k-1}^{(i,j)} x_{k|k-1}^{(i,j)}$
7:    **end if**
8: **end for**
9: $\hat{M}_{k|k-1} \leftarrow \ell$

**Sensor-control:**
10: **for all** $u_k \in \mathbb{U}$ **do**,         ▷ Constructing the PIMS:
11:    $\mathring{Z} \leftarrow \emptyset$.
12:    **for** $\ell = 1, \hat{M}_{k|k-1}$ **do**
13:       $\xi \leftarrow \arg\max_z g(z|\hat{x}_{k|k-1}^{(\ell)})$
14:       $\mathring{Z} \leftarrow \mathring{Z} \cup \{\xi\}$
15:    **end for**
                                                 ▷ Update for sensor control:
16: Use the PIMS $\mathring{Z}$ in (7)-(8) to update the multi-Bernoulli distribution parameters.
                                                 ▷ Computing the cost:
17: $\varepsilon_{|X|}^2(u_k) \leftarrow \left[\sum_{i=1}^{M_{k,u_k}} r_{k,u_k}^{(i)}(1-r_{k,u_k}^{(i)})\right]/\frac{M_{k,u_k}}{4}$.
18: $\sigma_{x(i)}^2(u_k) \leftarrow \sum_{j=1}^{L_{k,u_k}^{(i)}} w_{k,u_k}^{(i,j)} \mathsf{x}_{k,u_k}^{(i,j)2} - \left(\sum_{j=1}^{L_{k,u_k}^{(i)}} w_{k,u_k}^{(i,j)} \mathsf{x}_{k,u_k}^{(i,j)}\right)^2$
19: $\sigma_{y(i)}^2(u_k) \leftarrow \sum_{j=1}^{L_{k,u_k}^{(i)}} w_{k,u_k}^{(i,j)} \mathsf{y}_{k,u_k}^{(i,j)2} - \left(\sum_{j=1}^{L_{k,u_k}^{(i)}} w_{k,u_k}^{(i,j)} \mathsf{y}_{k,u_k}^{(i,j)}\right)^2$
20: $\max\{\sigma_{x(i)}^2(u_k)\} \leftarrow \frac{1}{L_{k,u_k}^{(i)}}(1-\frac{1}{L_{k,u_k}^{(i)}})\sum_{j=1}^{L_{k,u_k}^{(i)}} \mathsf{x}_{k,u_k}^{(i,j)2}$
21: $\max\{\sigma_{y(i)}^2(u_k)\} \leftarrow \frac{1}{L_{k,u_k}^{(i)}}(1-\frac{1}{L_{k,u_k}^{(i)}})\sum_{j=1}^{L_{k,u_k}^{(i)}} \mathsf{y}_{k,u_k}^{(i,j)2}$
22: $\varsigma_{x(i)}^2(u_k) \leftarrow \frac{\sigma_{x(i)}^2(u_k)\ \sigma_{y(i)}^2(u_k)}{\max\{\sigma_{x(i)}^2(u_k)\}\ \max\{\sigma_{y(i)}^2(u_k)\}}$
23: $\varepsilon_X^2(u_k) \leftarrow \left[\sum_{i=1}^{M_{k,u_k}} r_{k,u_k}^{(i)} \varsigma_{x(i)}^2(u_k)\right] / \sum_{i=1}^{M_{k,u_k}} r_{k,u_k}^{(i)}$
24: $\mathcal{V}(u;X_k) \leftarrow \eta\ \varepsilon_{|X|}^2(u) + (1-\eta)\ \varepsilon_X^2(u)$,
25: **end for**
26: $\overset{\star}{u} \leftarrow \arg\min_u \mathcal{V}(u;X_k)$

**Measurement:**
27: Apply the control command $\overset{\star}{u}$ to change the sensor state.
28: Read the actual measurement set $Z_k$.

**Update:**
29: Use the set $Z_k$ as measurement set in equations (7)-(8) and compute the updated multi-Bernoulli parameters.
30: Prune and merge the updated Bernoulli components.   ▷ More details in [43].

### A. Case study 1

A controllable range-only sensor is used to track multiple targets in a surveillance application. Each single target state is comprised of location and velocity components in x and y directions, denoted by $[\mathsf{x}\ \mathsf{y}\ \dot{\mathsf{x}}\ \dot{\mathsf{y}}]^\top$. In the particular scenario borrowed from [14], [35], there are five targets. The sensor location is denoted by $\mathbf{s} = [\mathsf{x}_s\ \mathsf{y}_s]^\top$. The sensor can detect an object in location $\mathbf{o} = [\mathsf{x}_o\ \mathsf{y}_o]^\top$ with the following location-dependent probability:

$$p_D(\mathbf{s},\mathbf{o}) = \begin{cases} 1, & \text{if } \|\mathbf{o}-\mathbf{s}\| \leq R_0 \\ \max\{0, 1-\mathfrak{h}(\|\mathbf{o}-\mathbf{s}\|-R_0)\}. & \text{otherwise} \end{cases} \quad (18)$$

To make the results comparable, we used the same parameters used in [14], [35]. Those are as follows:
- surveillance area: $1000\,\text{m} \times 1000\,\text{m}$,
- $R_0 = 320\,\text{m}$,
- $\mathfrak{h} = 0.00025\,\text{m}^{-1}$,
- measurement model: $z = \|\mathbf{o}-\mathbf{s}\| + e$ where $e \sim \mathcal{N}(0,\sigma^2)$, $\sigma = \sigma_0 + \beta\|\mathbf{o}-\mathbf{s}\|^2$, $\sigma_0 = 1\,\text{m}$ and $\beta = 5\times 10^{-5}\,\text{m}^{-1}$.

The Poisson RFS for clutter has the intensity $\nu(z) = \lambda_c\ c(z)$ where $c(z) = U[0,1000\sqrt{2}]$, and $\lambda_c = 0.5$.

Overall, there are five objects in the surveillance area, positioned relatively close to each other. Their initial state vectors are: $[800\ 600\ 1\ 0]^\top$, $[650\ 500\ 0.3\ 0.6]^\top$, $[620\ 700\ 0.25\ 0.45]^\top$, $[750\ 800\ 0\ 0.6]^\top$, and $[700\ 700\ 0.2\ 0.6]^\top$, where the units of x and y are meters and $\dot{\mathsf{x}}$ and $\dot{\mathsf{y}}$ are m/s. The objects move according to the constant velocity model [9], [14], [22], [35].

The controllable mobile sensor initially enters the surveillance area at position $\mathbf{s} = \begin{bmatrix}10\,\text{m} & 10\,\text{m}\end{bmatrix}^\top$. Other parameters such as the dynamic parameters of the motion model, probability of survival, target birth model, finite set of admissible control commands, and relative number of particles are also borrowed from [14], [35]. As it is also noted in [35], the range dependent sensor noise and moderate rates of clutter and miss-detections make this a very challenging case of sensor-control for multi-target estimation in which many state-of-the-art techniques would fail.

Figure 1 shows the initial and final locations of the five targets in this scenario, and demonstrates how the sensor location is controlled towards locations with better accuracy and detection rate by PEECS method (Algorithm 1). We have compared PEECS sensor-control with recent PHD-based sensor control methods in terms of their OSPA errors. In addition, we compared our method with the multi-Bernoulli-based sensor-control method of Hoang et al. [14] which is more similar. Figure 1 shows the controlled sensor locations of this method compared to PEECS sensor-control method. We will further elaborate on this comparison later.

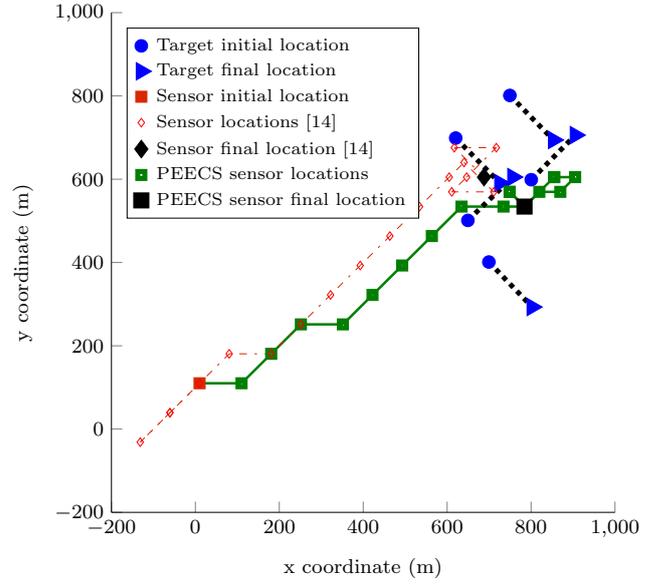

Fig. 1. Sensor locations for PEECS cost function and MAP cardinality variance cost function [14] during $k = 1, \ldots, 20$ for range-only measurement.

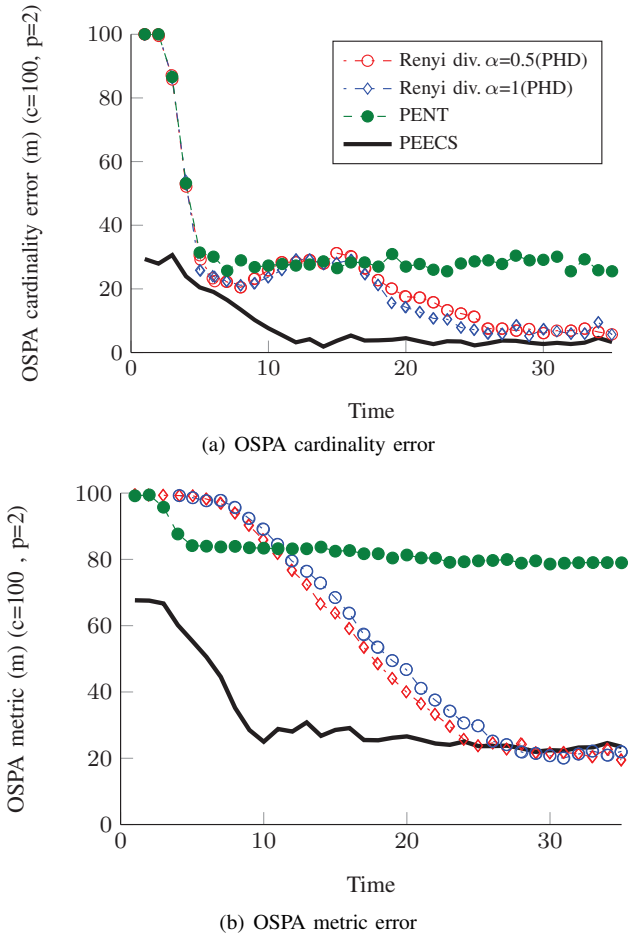

(a) OSPA cardinality error

(b) OSPA metric error

Fig. 2. Estimation errors of the PEECS sensor-control method compared to the PHD-based sensor-control.

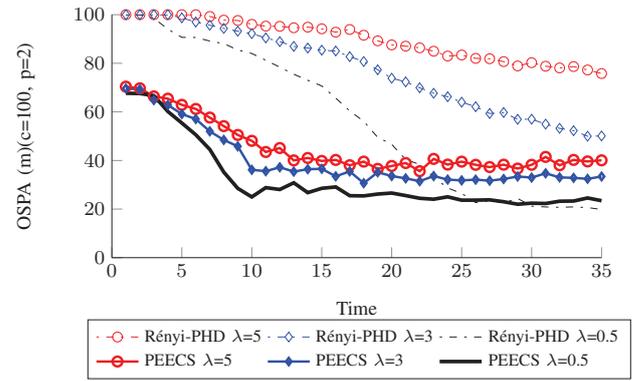

Fig. 3. Estimation errors with different clutter rate of our sensor-control method compared to the PHD-based sensor-control.

We computed the localization, cardinality and OSPA errors [36] at time steps $k = 1, \ldots, 35$. The estimation errors for three sensor-control methods using different reward/cost functions (PENT [24], Rényi divergence [35] and PEECS) are shown in Fig. 2 for the case of clutter rate $\lambda_c = 0.5$. Note that OSPA errors for Rényi divergence are computed and plotted for $\alpha = 0.5$ and $\alpha = 1$. We observe that during the initial time steps, the OSPA error returned by the PEECS sensor-control is significantly smaller than (about half of) the error of the two PHD-based sensor-control methods reported in [35] which have been demonstrated to outperform PENT and other existing methods. However, as it is reported in [35], the reason of poor performance of PENT lies in the fact that its objective function is designed for sensor-control problems in which the sensor has finite field of view, which is not the case in this scenario.

Although PEECS and PHD-based methods converge to similar error values, the initially smaller error of the PEECS sensor-control demonstrates that the sensor is guided towards its optimum location faster and arrives there much earlier than the PHD-based methods that work based on maximizing Rényi divergence. A video of the live simulation run is attached as supplemental material. It is observed from the run that after the first 13 time steps, our sensor-control drives the sensor to a close vicinity of the center point of the five targets, and during the rest of the simulation, the sensor stays in that region with minor movements. This observation is consistent with the OSPA errors of PEECS shown in figures 2(a) and 2(b) converging to its minimum at about time step 13.

To investigate the robustness of PEECS sensor-control to increased levels of clutter, we compared its OSPA errors with the PHD-based sensor-control method reported in [35] for different clutter rates of $\lambda_c = 0.5, 3, 5$. The errors were averaged over 200 Monte Carlo runs of each method and are shown in Fig. 3. It is observed that by increasing the clutter rate $\lambda_c$ from 0.5 to 3 then to 5, our proposed method returns OSPA errors that are generally smaller than the errors returned by the PHD-based sensor-control. We also note that the presented OSPA errors in the paper only provide a broad indication of sensor-control performance, and such results are scenario dependent.

The most significant advantage of PEECS sensor-control method is its low computational cost. Table I shows the computation times of PEECS sensor-control compared with the PHD-based sensor-control method of [35], averaged over 200 Monte Carlo runs, for different clutter rates.[5] We observe that in average, PEECS sensor-control runs at least 4 times faster than competing methods. Note that the result of the computation times should only be considered as a broad indication of sensor-control performance, since these results are scenario and code dependent. We also note that PEECS does not consider possible constraints on the sensor field of view and needs to be revised in presence of such constraints. Comparison of the computational cost of the revised version of PEECS with methods like PENT [27] (that take such limits into consideration) is beyond the scope of this paper and left

---

[5]Both methods were coded in MATLAB and run on the same machine. For the PHD-based sensor-control, we used the code provided by the authors of [35].

TABLE I
COMPARISON OF COMPUTATION TIMES (SECONDS)

| Clutter rate ($\lambda_c$) | 0.5 | 3.0 | 5.0 |
|---|---|---|---|
| PEECS | 0.310 | 0.333 | 0.373 |
| PHD-based method [35] | 1.336 | 1.631 | 1.973 |

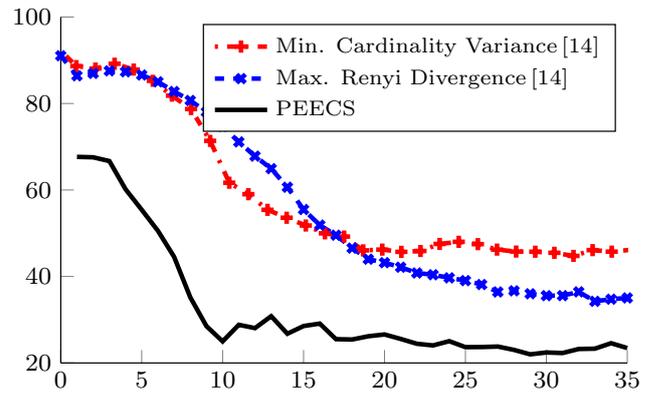

Fig. 4. The comparative results for PEECS method and the methods reported in [14]. In this case, clutter rate is burrowed from [14] to be $\lambda_c = 0.5$.

for future research.

The lower computational cost of PEECS sensor-control is mainly due to the fact that it is designed to work with the multi-Bernoulli filter. Extraction of cardinality and single-object state estimates in a multi-Bernoulli scheme is straightforward compared to PHD and CPHD methods. In the latter, the particles approximating the intensity function need to be grouped via a clustering algorithm. However, no such clustering step is required by multi-Bernoulli methods, hence substantial savings in computation is made particularly in applications that involve numerous objects. In such applications, the total number of particles, representing Monte Carlo approximations of the intensity function, is required to be very large otherwise substantial amount of information would be lost during the clustering step. The need for a large number of particles would impose heavy computation burden in those filtering schemes.

In Fig. 4, OSPA error of PEECS sensor-control method is compared with the method reported in [12], [14]. Hoang et al. [14] used multi-Bernoulli filter in conjunction with Rényi divergence as a reward function. We observe that PEECS sensor-control performs better than the Rényi divergence-based sensor control method proposed in [12], [14] in terms of OSPA error. This is mainly because PEECS cost function is structurally related to the error terms computed in OSPA. We note that the superior performance of PEECS in terms of OSPA error may not be necessarily valid in terms of information gain which is the major focus of Rényi divergence-based sensor control methods.

Hoang and Vo [14] also proposed a multi-Bernoulli sensor-control method in which the variance of cardinality around its MAP estimate (termed "MAP cardinality variance") is used as the cost function. This cost function is similar to PEECS. However, in PEECS, the variance term is the statistical variance (around the mean, not around the MAP estimate) of cardinality, and more importantly, the cost is penalized due to errors not only in cardinality estimates but also in state estimates. As it is shown in Fig. 4, our method outperforms the method reported in [12], [14] in terms of OSPA errors..

The controlled sensor locations resulted by PEECS and the method reported in [14] (with MAP cardinality variance as the cost function) are shown in Fig. 1. We note that in the case studies discussed in this paper, some measurement parameters depend on the sensor-target distance in such a way that a shorter sensor-target distance generally leads to more accurate measurements. To be more precise, based on the measurement models in case study 1, with a shorter sensor-target distance, higher probability of detection and smaller measurement noise are achieved. This seems to be a commonly accepted assumption as it is reported in many similar works on sensor control such as [14], [34], [35]. With such a distance-dependent measurement model, an effective sensor control method is expected to guide the sensor towards the center of the targets, and maintain it in the vicinity of the center as the targets move. Figure 1 clearly shows that PEECS directly moves the sensor towards the center of the targets and maintains it in the center as the targets move. This is while the sensor trajectory created by the competing method of [14] seems to be deviated from the path to the center of the targets and not to converge and remain in the center as the targets move. Hence, PEECS leads to the significantly improved OSPA errors shown in Fig. 4.

As stated by Hoang and Vo [14], "the lack of observability of the full states, as we now have range-only measurements," causes the poor performance of their proposed sensor control, especially when only the variance of cardinality around its MAP estimate is optimized. If the objective function "accounts for both localization and cardinality criteria", which is indirectly the case when Rényi divergence is chosen as the objective function, sensor control is expected to result in "lower error" [14]. We note that both cardinality and state estimation errors are explicitly considered in PEECS objective function.

### B. Case study 2

To demonstrate the performance of our method with measurements that guarantee full observability of the targets, and to investigate the particular effect of the localization error term in the proposed cost function, we also designed a simulation involving a more complex scenario than case study 1. In this scenario, we chose the non-linear nearly-constant turn model employed in [43]. In this model, each single target state $x = [\bar{x}^\top \; \omega]^\top$ is comprised of location and velocity in Cartesian coordinates, denoted by $\bar{x} = [\text{x} \; \text{y} \; \dot{\text{x}} \; \dot{\text{y}}]^\top$ and turning rate, denoted by $\omega$. The state dynamics are given by:

$$\bar{x}_k = F(\omega_{k-1})\bar{x}_{k-1} + G\epsilon_{k-1},$$
$$\omega_k = \omega_{k-1} + T\gamma_{k-1},$$

where
$$F(\omega) = \begin{bmatrix} 1 & 0 & \frac{\sin\omega T}{\omega} & -\frac{1-\cos\omega T}{\omega} \\ 0 & 1 & \frac{1-\cos\omega T}{\omega} & \frac{\sin\omega T}{\omega} \\ 0 & 0 & \cos\omega T & -\sin\omega T \\ 0 & 0 & \sin\omega T & \cos\omega T \end{bmatrix}, G = \begin{bmatrix} \frac{T^2}{2} & 0 \\ 0 & \frac{T^2}{2} \\ T & 0 \\ 0 & T \end{bmatrix},$$

$T = 1\,\text{s}$, $\epsilon_{k-1} \sim \mathcal{N}(\cdot; 0, \sigma_\epsilon^2 I)$, $\sigma_\epsilon = 15\,\text{m/s}^2$, and $\gamma_{k-1} \sim \mathcal{N}(\cdot; 0, \sigma_\gamma^2 I)$, $\sigma_\gamma = (\pi/180)\,\text{rad/s}$. The birth RFS is a multi-Bernoulli with density $\pi_\Gamma = \{(r_\Gamma^{(i)}, p_\Gamma^{(i)})\}_{i=1}^4$ where $r_\Gamma^{(1)} = r_\Gamma^{(2)} = 0.02$, $r_\Gamma^{(3)} = r_\Gamma^{(4)} = 0.03$ and $p_\Gamma^{(i)}(x) = \mathcal{N}(x; m_\gamma^{(i)}, P_\gamma)$ where

$$\begin{aligned} m_\gamma^{(1)} &= [-1500\ 0\ 250\ 0\ 0]^\top, \\ m_\gamma^{(2)} &= [-250\ 0\ 1000\ 0\ 0]^\top, \\ m_\gamma^{(3)} &= [250\ 0\ 750\ 0\ 0]^\top, \\ m_\gamma^{(4)} &= [1000\ 0\ 1500\ 0\ 0]^\top, \\ P_\gamma &= \text{diag}(50^2, 50^2, 50^2, 50^2, (6 \times \tfrac{\pi}{180})^2). \end{aligned}$$

Probability of survival, detection probability, initial sensor location and clutter rate are similar to the previous case study. In this case study, in addition to range, bearing information is also available and the observation model consists of noisy bearing and range measurements:

$$z_k = \begin{bmatrix} \arctan(\tfrac{x_k}{y_k}) & \sqrt{x_k^2 + y_k^2} \end{bmatrix}^\top + \zeta_k,$$

where $\zeta_k \sim \mathcal{N}(\cdot; 0, R_k)$ is the measurement noise with covariance $R_k = \text{diag}(\sigma_\theta^2, \sigma_r^2)$ in which the scales of range and bearing noise are $\sigma_\theta = (\pi/180)\,\text{rad}$ and $\sigma_r = 5\,\text{m}$. The clutter RFS followed the uniform Poisson model over the surveillance region $[-\pi/2, \pi/2]\,\text{rad} \times [0, 2000]\,\text{m}$, with intensity of $1.6 \times 10^{-3}\,(\text{rad\,m})^{-1}$ which is equivalent to a clutter rate of $\lambda_c = 10$.

As it is shown in Fig. 5, the sensor starts moving toward the objects and remains between those. We ran PEECS sensor-control algorithm for both $\eta = 1$ and $\eta = 0.5$. When the parameter $\eta$ equals 1, the cost function includes only the cardinality error term. With $\eta = 0.5$, both cardinality error and object state estimation error are equally weighted within the cost. We recorded the OSPA errors in both cases at time steps $k = 1, \ldots, 50$. The results are plotted in Fig. 6. this figure shows that although the targets maneuver in long paths (compared to the case study 1 borrowed from [35]), they are tracked with reasonably low error, indicating the reliability of PEECS. Comparison of OSPA errors in cases where $\eta = 1$ or $\eta = 0.5$ also demonstrates the benefit gained in terms of total error when both cardinality and state estimation errors were considered within the cost function, as it happened in PEECS with $\eta = 0.5$.

## VII. CONCLUSION

A sensor-control solution was proposed to be employed within a multi-Bernoulli multi-target filter. In this method, at each step, the next sensor control command is chosen by minimizing a new task-driven cost function called "Posterior Expected Error of Cardinality and States" (PEECS). The PEECS cost associated with each control command is defined as a linear combination of the normalized errors in cardinality estimate of the multi-target random finite set and the normalized error of localization of the elements in the multi-target RFS. Simulation results involving two challenging multi-target estimation and sensor-control scenarios, demonstrated that PEECS sensor-control can return multi-object state estimation accuracy and clutter tolerance that are similar to or better than competing methods, and generally performs faster than the state-of-the-art.

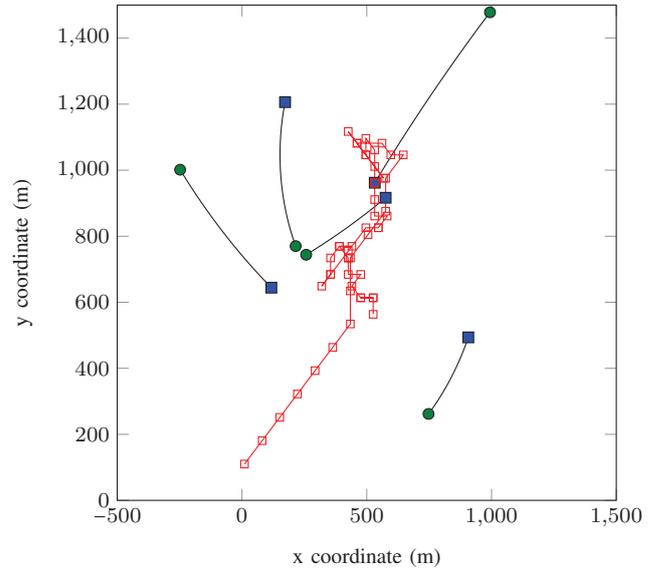

Fig. 5. Sensor and target locations in case study 2, during $k = 1, \ldots, 50$.

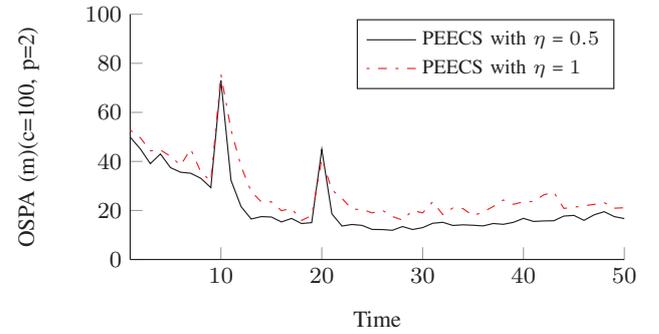

Fig. 6. OSPA errors returned by PEECS in case study 2.

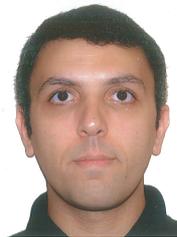

**Amirali K. Gostar** received his Bachelors degree in Electrical Engineering in 2000. He continued his study in Philosophy of Science and received his Masters degree in 2006. He is currently a PhD candidate at the School of Aerospace, Mechanical and Manufacturing Engineering, RMIT University. His research interests include sensor management, data fusion, multi-target tracking and simulation methods.

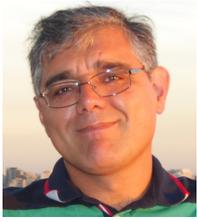

**Reza Hoseinnezhad** received his B.Sc., M.Sc. and Ph.D. degrees in Electronic, Control and Electrical Engineering all from the University of Tehran, Iran, in 1994, 1996 and 2002, respectively. Since 2002, he has held various academic positions at the University of Tehran, Swinburne University of Technology, the University of Melbourne and RMIT University. He is currently a senior lecturer with the School of Aerospace, Mechanical and Manufacturing Engineering, RMIT University, Victoria, Australia. His research is currently focused on development of robust estimation and visual tracking methods in a point process framework.

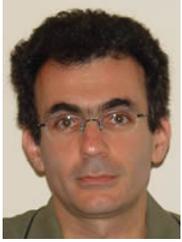

**Alireza Bab-Hadiashar** received his BSc degree in mechanical engineering from Tehran University (1988), MES degree in mechanical and mechatronics engineering from The University of Sydney (1993) and the PhD degree in computer vision from Monash University (1997). His main research interest is the development of robust estimation and segmentation techniques for computer vision applications. He is currently a professor in the School of Aerospace, Mechanical and Manufacturing Engineering at RMIT University, Australia and is a senior member of the IEEE.